\documentclass[10pt, a4paper]{article}
\usepackage{epsfig}
\usepackage{graphicx}               
\usepackage{amsmath}                
\usepackage{multirow}               
\usepackage{fancyhdr}
\usepackage{here}
\usepackage{comment}
\usepackage[dvipdfmx]{hyperref}

\begin{document}

\pagestyle{fancy}

\title{\LARGE\bf Quark charge identification for $e^{+}e^{-}$ to $q\bar{q}$ study}
\author{
  {\large Y.~Uesugi$^1$\thanks{Presenter. Talk presented at the International Workshop on Future Linear Colliders (LCWS2018), Arlington, Texas,22-26 October 2018. }, H.~Yamashiro$^1$, T.~Suehara$^1$, T.~Yoshioka$^2$, K.~Kawagoe$^1$} \\ \\
  $^1$Department of Physics, Faculty of Science, Kyushu University \\
  $^2$Research Center for Advanced Particle Physics, Kyushu University \\
  744 Motooka, Nishi-ku, Fukuoka, 819-0395 Japan
 }
\date{}

\maketitle
\thispagestyle{fancy}
\setcounter{page}{1} 

\begin{abstract}
The process $e^{+}e^{-} \to q\bar{q}$ plays an important role in electroweak precision measurements. We are studying this process with ILD full simulation.
The key for the reconstruction of the quark pair final states is quark charge identification (ID). We report the progress of charge ID study in detail. In particular, we investigate the performance of the charge ID for each decay mode of the heavy hadrons to know the possibilities of improvements of the charge ID.
\end{abstract}

\section{Introduction}

Two-quark final states in the high energy $e^+e^-$ collisions are important for precise measurements of the electroweak interaction. These simple processes have low background and the uncertainty has a little effect on QED calculation. We are studying the $e^+e^- \to q\bar{q}$ final states in the International Linear Collider (ILC) for a probe to new physics.
The angular distribution with respect to the beam axis is used to calculate the sensitivity to new physics beyond the Standard Model\cite{Yamashiro-proc}. Identification of quark charge is necessary to separate the angular distribution of positive and negative quarks.
The efficiency of quark charge ID in the previous study\cite{Yamashiro} was about 60\%, which causes significant performance degradation due to the misidentification of the quark charge.
For the performance improvement, we investigate the performance of charge ID of the current software
for each decay mode of $b$ hadrons.

\section{Simulation condition}

 We utilized ILCSoft\cite{ILCSoft} version v01-16 for this study. The event samples of $e^{+}e^{-} \to b\bar{b}$ was generated in center of mass energy of 250 GeV by WHIZARD 1.95\cite{Whizard}. The full Monte Carlo simulation (MC) was done with Mokka based on Geant4 framework with the reference geometry of the International Large Detector (ILD) concept used in the studies of Detailed Baseline Design report\cite{DBD}, ILD\_v1\_o5 model. The model includes silicon pixel and strip detectors, a time projection chamber, precisely segmented electromagnetic and hadron calorimeters (ECAL and HCAL) and a 3.5 Tesla solenoid magnet. 
Event reconstruction was done with Marlin processors, including tracking and particle flow reconstruction by PandoraPFA algorithm\cite{pfa} to obtain track-cluster matching. The reconstructed particles were clustered to two jets with Durham\cite{Durham} algorithm.

\section{Vertex finder and quark charge identification}

The key feature to reconstruct the quark charge is a vertex finder. We used LCFIPlus\cite{LCFIPlus} to reconstruct vertices. In the LCFIPlus, all tracks are firstly processed with a primary vertex finder based on tear-down technique, with the beam constraint to remove most of tracks consistent to come from the interaction point.
Then remaining tracks are processed with a secondary vertex finder based on build-up technique.
It does not restrict the number of vertices per jet to reconstuct, but after the jet clustering
it combines the vertices if there are more than two vertices in the jet with refitting vertex positions.
In addition, it uses tracks which are consistent to cross the line of the primary vertex and secondary vertices
to formulate additional pseudo-vertices, which aims to recover vertices having only one track.
Overall fraction of having two vertices in $b$ quark jet with the current implementation is
around 40\%, and having one or more vertices is around 80\%.
In the following discussion, we treat two vertices separately if we obtain two vertices in a jet,
and treat the vertex as combined vertices of $b$ and $c$ decay if we obtain only one vertex.

To separate $b$ and $\bar{b}$, the charges of the second and third vertex are calculated as the sum of track charge associated to the vertices.

There are four ways of obtaining quark charge as follows:
\begin{itemize}
  \item jet charge (charge sum of all tracks in the jet) (${\rm\Sigma_{all}^{jet}}$) 
  \item vertex charge
  \begin{itemize}
   \item charge sum of tracks associated to the second vertex ($\rm\Sigma_{vtx}^{2nd}$)
   \item charge sum of tracks associated to the third vertex (if found) ($\rm\Sigma_{vtx}^{3rd}$)
   \item charge sum of tracks associated to the second and third vertex ($\rm\Sigma_{vtx}^{2nd 3rd}$)
  \end{itemize}
\end{itemize}

For jets with two vertices found, we can use $\rm\Sigma_{vtx}^{2nd}$ and $\rm\Sigma_{vtx}^{3rd}$ to separate $b$ decay, but for jets with only one vertex found, we can only use $\rm\Sigma_{vtx}^{2nd 3rd}$.
For jets without vertices found, we have to use ${\rm\Sigma_{all}^{jet}}$ but this is not discussed in this study.


\section{Decay modes of B mesons}

The purpose of charge ID is to distinguish jets from $b$ quarks and jets from $\bar{b}$ quarks.
Each $b$ or $\bar{b}$ quark formulates a $b$ hadron after fragmentation,
which is usually in the core of the jet. Table \ref{tab:Bhadrons} shows $b$ hadrons obtained from
$b$ and $\bar{b}$ quarks separately. The charge of the $b$ quark is closely related
to the quark constituents in the final state $b$ hadrons, for example $B^-$ and $\bar{B^0}$
only come from $b$ quark and not from $\bar{b}$ quark, and $B^+$ and ${B^0}$ only come from $\bar{b}$ quark
and not from $b$ quark if we ignore quark-antiquark oscillation.
If we can separate $B^0$ and $\bar{B^0}$, this can enhance the charge ID performance significantly, compared to just identify the charge of the $b$ hadrons. This can be realized by observing decay of $B$ mesons.
Here we only focus on the $B^+$, $B^0$ and those antiparticles, however, similar discussion can be done with other decay modes as well.

\begin{table}[htbp]
 \begin{center}
  \begin{tabular}{|c|c||c|c|} \hline
   \multicolumn{2}{|c||}{$b$} & \multicolumn{2}{c|}{$\bar{b}$}\\ \hline
    \multicolumn{1}{|c|}{-1}
      & \multicolumn{2}{c|}{0} & \multicolumn{1}{c|}{+1}\\  \hline
    $B^{-}  (\bar{u}b)$ & $\bar{B}^{0}  (\bar{d}b)$ & $B^{0}  (d\bar{b})$ & $B^{+}  (u\bar{b})$ \\ 
                                &                                       &          42.2\%          &               41.9\%     \\ \hline
    $B^{-}_{c}  (\bar{c}b)$ & $\bar{B}^{0}_{s}  (\bar{s}b)$ & $B^{0}_{s}  (s\bar{b})$ & $B^{+}_{c}  (c\bar{b})$ \\ 
                                     &                                            &              8.0\%            &                                    \\ \hline
    $\Xi^{-}_{b}  (dsb)$ & $\Lambda^{0}_{b}  (udb)$ & $\bar{\Lambda}^{0}_{b}  (\bar{u}\bar{d}\bar{b})$ & $\Xi^{+}_{b}  (\bar{d}\bar{s}\bar{b})$ \\
                                 &                                        &          6.4\%                                                        &               0.61\%                               \\ \hline
    $\Omega^{-}_{b}  (ssb)$ & $\Xi^{0}_{b}  (usb)$ & $\bar{\Xi}^{0}_{b}  (\bar{u}\bar{s}\bar{b})$ & $\Omega^{+}_{b}  (\bar{s}\bar{s}\bar{b})$ \\
                                &                                       &          0.59\%          &               0.010\%      \\ \hline
  \end{tabular}
  \caption{List of semistable $b$ hadrons produced from $b$ and $\bar{b}$ quarks. The production ratio, obtained from MC information, is also shown for $\bar{b}$, which ignores $b$-$\bar{b}$ oscillation. Production of charmed $b$ hadrons ($B_c$), which usually decay to lighter $b$ hadrons initially, are included in the fraction of lighter $b$ hadrons.}
  \label{tab:Bhadrons}
  \end{center}
\end{table}

Table \ref{tab:br} shows the dominant decay modes of $B^+$ and $B^0$ mesons.
As shown in the tables, there is some discrepancy of the branching ratios between
PDG values and those obtained from the MC samples, which may be due to the
$b$-$\bar{b}$ oscillation. For the $B^+$ decay, the dominant decay mode is
$B^+ \to \bar{D^0}X$, which gives positive vertex on the decay of $B^+$ and
neutral vertex on the subsequent decay of $\bar{D^0}$.
In the case of $B^0$ decay, there are two dominant decay modes, $B^0 \to \bar{D^0}X$
and $B^0 \to D^-X$. For the latter decay, the decay vertex of $B^0$ should be positive
 and the subsequent $D^-$ decay should be negative, which should have separation
 power from $\bar{B^0}$. For this separation, separation of second and third vertices is critical.

\begin{table}[htbp]
 \begin{minipage}{0.5\hsize}
 \begin{center}
  \begin{tabular}{|c|c|c|} \hline
    Decay modes & BR in PDG & BR in MC \\ \hline      
    $B^{+} \to \bar{D}^{0}X$ & 79\% & 70.51\% \\ \hline
    $B^{+} \to D^{-}X$ & 9.9\% & 9.81\% \\ \hline
    $B^{+} \to D^{0}X$ & 8.6\% & 4.21\% \\ \hline
    $B^{+} \to D^{+}_{s}X$ & 7.9\% & 4.80\% \\ \hline
    $B^{+} \to D^{+}X$ & 2.5\% & 1.65\% \\ \hline
     \end{tabular}
  \end{center}
  \end{minipage}
  \begin{minipage}{0.5\hsize}
 \begin{center}
  \begin{tabular}{|c|c|c|} \hline
    Decay modes & BR in PDG & BR in MC \\ \hline      
    $B^{0} \to \bar{D}^{0}X$ & 47.4\% & 40.24\% \\ \hline
    $B^{0} \to D^{-}X$ & 36.9\% & 27.59\% \\ \hline
    $B^{0} \to D^{+}_{s}X$ & 10.3\% & 4.21\% \\ \hline
    $B^{0} \to D^{0}X$ & 8.1\% & 11.55\% \\ \hline
    $B^{0} \to D^{+}X$ & $<$3.9\% & 6.91\% \\ \hline
    
  \end{tabular}
  \end{center}
  \end{minipage}
  \caption{Branching ratios (BR) of $B^+$ (left) and $B^0$ (right) mesons. BR in PDG is from \cite{PDG}, and BR in MC is from the event sample.}
  \label{tab:br}
\end{table}

\section{Current performance of charge ID}

The performance of charge ID with LCFIPlus is checked with $B^+$ and $B^0$ data.
After separation of the decay mode with MC information, $b$ and $\bar{b}$ quarks are
assigned to jets using MC-track matching. The reconstructed vertices are examined and
$\rm\Sigma_{vtx}^{2nd}$, $\rm\Sigma_{vtx}^{3rd}$ and $\rm\Sigma_{vtx}^{2nd 3rd}$
are calculated for each jet.

Table \ref{tab:3} shows those observables of $B^+$ decays, categorized by the number of reconstructed vertices and $B^+$ decay modes (with only first and second dominant decay).
For the events with 1 vertex found, it shows that positive $\rm\Sigma_{vtx}^{2nd 3rd}$
is much more than negative charge, which proves that charge ID is possible.
However, there is significant amount of ``neutral" vertex, which limits the charge ID performance.
For the events with 2 vertices found, the positive $\rm\Sigma_{vtx}^{2nd}$ dominates
more than with 1 vertex case, thus gives better performance of charge ID.
For the special case of $B^{+} \to \bar{D}^{-}X$, $\rm\Sigma_{vtx}^{2nd}$ should be +2 and
$\rm\Sigma_{vtx}^{3rd}$ should be -1, which is much easier to identify.

\begin{table}[htbp]
 \begin{center}
 \begin{tabular}{|c|c|c|c|c|c|c|c|c|c|}
\hline
particle                & \multicolumn{3}{c|}{$B^{+}$(all decay)}                     & \multicolumn{3}{c|}{$B^{+} \to \bar{D}^{0}X$(79\%)} & \multicolumn{3}{c|}{$B^{+} \to \bar{D}^{-}X$(9.9\%)}                     \\ \hline
\multirow{2}{*}{number of vertex} & \multirow{2}{*}{1} & \multicolumn{2}{c|}{2} & \multirow{2}{*}{1} & \multicolumn{2}{c|}{2} & \multirow{2}{*}{1} & \multicolumn{2}{c|}{2} \\ \cline{3-4} \cline{6-7} \cline{9-10} 
                        &                    & 2nd        & 3rd       &                    & 2nd        & 3rd      &                    & 2nd        & 3rd \\ \hline
    symbol & $\rm\Sigma_{vtx}^{2nd 3rd}$ & $\rm\Sigma_{vtx}^{2nd}$ & $\rm\Sigma_{vtx}^{3rd}$ & $\rm\Sigma_{vtx}^{2nd 3rd}$ & $\rm\Sigma_{vtx}^{2nd}$ & $\rm\Sigma_{vtx}^{3rd}$ & $\rm\Sigma_{vtx}^{2nd 3rd}$ & $\rm\Sigma_{vtx}^{2nd}$ & $\rm\Sigma_{vtx}^{3rd}$ \\ \hline
    charge $<0$ & 8.67\% & 8.06\% & 22.8\% & 7.52\% & 7.21\% & 16.6\% & 17.2\% & 7.85\% & 67.6\% \\ \hline
    charge $=0$ & 35.5\% & 18.5\% & 53.0\% & 38.6\% & 18.1\% & 6.07\% & 20.0\% & 8.75\% & 20.0\% \\ \hline
    charge $>0$ & 55.7\% & 73.3\% & 24.0\% & 53.8\% & 74.6\% & 22.6\% & 62.7\% & 83.3\% & 12.3\% \\ \hline
  \end{tabular}
  \caption{Reconstructed charge of the vertices in $B^+$ decay. The left 3 columns show the charge of all decay modes, and the right 6 columns show the charge of two dominant decay modes.}
  \label{tab:3}
  \end{center}
\end{table}

Separation of $B^{0}$ and $\bar{B^0}$ is apparently more difficult
since the total charge of $B^0$ and subsequent charm hadron is neutral.
There are two dominant decay modes of $B^0$:
$B^{0} \to \bar{D}^{0}X$ and $B^{0} \to D^-X$.
The former is quite difficult to distinguish from $\bar{B^0} \to D^0$,
since both of second and third vertices are neutral. There may be some
possibility to identify kaons to check their charge, but this is beyond
the current study. In $B^{0} \to D^-X$ case we have a chance to
separate from $\bar{B^0}$ since the first $B^0$ vertex should be positive
and subsequent charm vertex should be negative, as shown in Table \ref{tab:7}.
Here the separation is nearly impossible with one vertex found, but with two vertices
there is a significant difference on the positive and negative fractions of
$\rm\Sigma_{vtx}^{2nd}$ and $\rm\Sigma_{vtx}^{3rd}$, which gives
separation power of $B^0$ and $\bar{B^0}$.

\begin{table}[htbp]
 \begin{center}
 \begin{tabular}{|c|c|c|c|}
\hline
particle                & \multicolumn{3}{c|}{$B^{0} \to \bar{D}^{-}X$(36.9\%)}  \\ \hline
\multirow{2}{*}{number of vertex} & \multirow{2}{*}{1} & \multicolumn{2}{c|}{2}  \\ \cline{3-4} 
                        &                    & 2nd        & 3rd         \\ \hline
    charge $<0$ & 29.5\% & 11.8\% & 72.3\% \\ \hline
    charge $=0$ & 33.2\% & 17.6\% & 17.5\% \\ \hline
    charge $>0$ & 37.1\% & 70.5\% & 10.0\% \\ \hline
  \end{tabular}
  \caption{Reconstructed charge of the vertices in $B^{0} \to \bar{D}^{-}X$ decay.}
  \label{tab:7}
  \end{center}
\end{table}

\section{Summary and prospects}

We investigated the performance of quark charge ID to be used in $e^{+}e^{-} \to q\bar{q}$ study.
By separating decay modes, we have a chance to separate $b$ and $\bar{b}$ by using
tracks from second and third vertices independently.
However, misidentification of the vertex charge is still to be improved. The main reason should be
tracks missed to be clustered into the vertex. We will investigate to recover those tracks by checking
behaviour of the vertex finder more precisely.
There is a trial of such a vertex recovery\cite{Sviatoslav} which should be revisited.

It is also important to increase the fraction of events which can find two vertices since
having two vertices gains the performance significantly.
We can also consider to use zero-vertex events by looking for secondary particles with
larger impact parameters or jet leptons.
Investigation of charm jets should also be done as a future plan.

\section*{Acknowledgements}

We appreciate the ILD software group for the support and producing event samples.
This work was supported by JSPS KAKENHI Grant Number 16H02176.


\begin{thebibliography}{99}
\bibitem{Yamashiro-proc} H. Yamashiro et al., Study of fermion pair productions at the ILC with center-of-mass energy of 250 GeV, Proc. LCWS2017, arXiv:1801.04671.
\bibitem{Yamashiro}H. Yamashiro, Master thesis at Kyushu University (2018),\\ \href{http://epp.phys.kyushu-u.ac.jp/thesis/2018MasterYamashiro.pdf}{http://epp.phys.kyushu-u.ac.jp/thesis/2018MasterYamashiro.pdf} (Japanese)
\bibitem{ILCSoft} \href{http://ilcsoft.desy.de/portal/}{http://ilcsoft.desy.de/portal/}
\bibitem{Whizard} W.~Kilian, T.~Ohi, J.~Reuter, Eur.~Phys.~J.~{\bf C71} (2011) 1742.
\bibitem{DBD} T.~Behnke {\it et al.}, The International Linear Collider Technical Design Report - Volume 4: Detectors. 2013, arXiv:1306.6329.
\bibitem{pfa} M.~A.~Thomson, Nucl.~Instrum.~Meth.~{\bf A611} (2009) 25-40.
\bibitem{Durham} S.~Catani {\it et al.}, Phys.~Lett.~{\bf B269} (1991) 432-438.
\bibitem{LCFIPlus} T.~Suehara, T.~Tanabe, Nucl.~Instrum.~Meth.~{\bf A808} (2016) 109-116.
\bibitem{PDG} M. Tanabashi {\it et al.} (Particle Data Group), \href{https://journals.aps.org/prd/abstract/10.1103/PhysRevD.98.030001}{Phys.~Rev.~{\bf D98} (2018) 030001}
\bibitem{Sviatoslav} S. Bilokin, R. P\"oschl, F. Richard, Measurement of $b$ quark EW couplings at ILC, arXiv:1709.04289.
\end{thebibliography}
\end{document}